\documentclass{emulateapj}
\slugcomment{}

\usepackage{graphicx}
\usepackage{natbib}

\def\deg      {{\ifmmode^\circ\else$^\circ$\fi}} 

\def\revised#1{{#1}}
 
\shorttitle{Galaxy clustering in the COSMOS field}
\shortauthors{McCracken et al.}

\begin{document}


 \title{The angular correlations of galaxies in the COSMOS field}


%
%
%

 \author{H. J. McCracken\altaffilmark{1,2},
J. A. Peacock\altaffilmark{3},
L. Guzzo\altaffilmark{4,5,6},
P. Capak\altaffilmark{7,8}
C. Porciani\altaffilmark{9},
N. Scoville\altaffilmark{7,8},
H. Aussel\altaffilmark{10},
A. Finoguenov\altaffilmark{11},
J. B. James\altaffilmark{3},
M. G. Kitzbichler\altaffilmark{5},
A. Koekemoer\altaffilmark{12},
A. Leauthaud\altaffilmark{13},
O. Le F\`{e}vre\altaffilmark{13},
R. Massey\altaffilmark{7},
Y. Mellier\altaffilmark{1,2},
B. Mobasher\altaffilmark{12},
P. Norberg\altaffilmark{3},
J. Rhodes\altaffilmark{7,14},
D. B. Sanders\altaffilmark{15},
S. S. Sasaki\altaffilmark{16,17},
Y. Taniguchi\altaffilmark{17},
D. J. Thompson\altaffilmark{18,19},
S. D. M. White\altaffilmark{5},
A. El-Zant\altaffilmark{20}
}


\altaffiltext{$\star$}{Based on observations with the NASA/ESA {\em
    Hubble Space Telescope},obtained at the Space Telescope Science
  Institute, which is operated by AURA Inc, under NASA contract NAS
  5-26555; also based on data collected at : the Subaru Telescope,
  which is operated by the National Astronomical Observatory of Japan;
  the XMM-Newton, an ESA science mission with instruments and
  contributions directly funded by ESA Member States and NASA; the
  European Southern Observatory under Large Program 175.A-0839, Chile;
  Kitt Peak National Observatory, Cerro Tololo Inter-American
  Observatory, and the National Optical Astronomy Observatory, which
  are operated by the Association of Universities for Research in
  Astronomy, Inc.  (AURA) under cooperative agreement with the National
  Science Foundation; the National Radio Astronomy Observatory which is
  a facility of the National Science Foundation operated under
  cooperative agreement by Associated Universities, Inc; and and the
  Canada-France-Hawaii Telescope with MegaPrime/MegaCam operated as a
  joint project by the CFHT Corporation, CEA/DAPNIA, the National
  Research Council of Canada, the Canadian Astronomy Data Centre, the
  Centre National de la Recherche Scientifique de France, TERAPIX and
  the University of Hawaii.}

\altaffiltext{1}{Institut d'Astrophysique de Paris, UMR7095 CNRS, Universit\`{e} Pierre et Marie Curie, 98 bis Boulevard Arago, 75014  Paris, France}
\altaffiltext{2}{Observatoire de Paris, LERMA, 61 Avenue de l'Observatoire, 75014 Paris, France}
\altaffiltext{3}{Institute for Astronomy, University of Edinburgh,  Royal Observatory, Blackford Hill, Edinburgh EH9 3HJ, U.K.}
\altaffiltext{4}{INAF-Osservatorio Astronomico di Brera, via Bianchi  46, I-23807 Merate (LC), Italy}
\altaffiltext{5}{Max-Planck-Institut f\"{u}r Astrophysik, D-85748 Garching bei M\"{u}nchen, Germany}
\altaffiltext{6}{Visiting Scientist, European Southern Observatory,Karl-Schwarzschild-Str. 2, D-85748 Garching, Germany}
\altaffiltext{7}{California Institute of Technology, MC 105-24, 1200 East California Boulevard, Pasadena, CA 91125}
\altaffiltext{8}{Visiting Astronomer, Univ. Hawaii, 2680 Woodlawn Dr.,  Honolulu, HI, 96822}
\altaffiltext{9}{Department of Physics, ETH Zurich, CH-8093 Zurich, Switzerland}
\altaffiltext{10}{Service d'Astrophysique, CEA/Saclay, 91191 Gif-sur-Yvette, France}
\altaffiltext{11}{Max-Planck-Institut f\"ur Extraterrestrische Physik, Giessenbachstra\ss e, 85748 Garching, Germany}

\altaffiltext{12}{Space Telescope Science Institute, 3700 San Martin Drive, Baltimore, MD 21218}
\altaffiltext{13}{Laboratoire d'Astrophysique de Marseille, BP 8, Traverse du Siphon, 13376 Marseille Cedex 12, France}

\altaffiltext{14}{Jet Propulsion Laboratory, Pasadena, CA 91109}

\altaffiltext{15}{Institute for Astronomy, 2680 Woodlawn Dr.,University of Hawaii, Honolulu, Hawaii, 96822}

\altaffiltext{16}{Astronomical Institute, Graduate School of Science, Tohoku University, Aramaki, Aoba, Sendai 980-8578, Japan}

\altaffiltext{17}{Physics Department, Graduate School of Science, Ehime University, 2-5 Bunkyou, Matuyama, 790-8577, Japan}

\altaffiltext{18}{Large Binocular Telescope Observatory, University of Arizona, 933 N. Cherry Ave.Tucson, AZ 85721-0065, USA}
\altaffiltext{19}{Caltech Optical Observatories, MS 320-47, California Institute of Technology, Pasadena, CA 91125}
\altaffiltext{20}{Canadian Institute for Theoretical Astrophysics, Mclennan Labs, University of Toronto, 60 St. George St, Room 1403, Toronto, ON M5S 3H8, Canada}

\begin{abstract}

  We present measurements of the two-point galaxy angular correlation
  function $w(\theta)$ in the COSMOS field. Independent determinations
  of $w(\theta)$ as a function of magnitude limit are presented for
  both the HST ACS catalog and also for the ground-based data from
  Subaru and the CFHT. Despite having significantly different masks,
  these three determinations agree well. At bright magnitudes ($I_{\rm
    AB}<22$), our data generally match very well with existing
  measurements and with mock catalogs based on semi-analytic galaxy
  formation calculations of \citet{2006astro.ph..9636K} from the
  Millennium Simulation. The exception is that our result is at the
  upper end of the expected cosmic variance scatter for $\theta > 10$
  arcmin, which we attribute to a particularly rich structure known to
  exist at $z\simeq 0.8$.  For fainter samples, however, the level of
  clustering is somewhat higher than reported by some previous
  studies: \revised {in all three catalogues} we find
  $w(\theta=1')\simeq 0.014$ at a median $I_{\rm AB}$ magnitude of
  24. At these very faintest magnitudes, our measurements agree well
  with the latest determinations from the Canada-France Legacy Survey.
  This level of clustering is approximately double what is predicted
  by the semi-analytic catalogs (at all angles).  The semi-analytic
  results allow an estimate of cosmic variance, which is too small to
  account for the discrepancy. We therefore conclude that the mean
  amplitude of clustering at this level is higher than previously
  estimated.

\end{abstract}

\section{Introduction}
The COSMOS field \citep{scoville_cosmos_overview} is the largest
contiguous multi-wavelength probe of the high-redshift galaxy
distribution, and a major task for the survey will be to extract
improved measurements of galaxy clustering at these early times. In
this initial paper, we will be concerned with the simplest of these
measures: the angular two-point correlation function, $w(\theta)$.
Demonstrating a robust measurement of this quantity is a minimum
requirement for verifying that the survey completeness is understood,
as a basis for future more elaborate analysis. The main aim of this
paper is therefore to present measurements of the two-point galaxy
clustering statistic on the COSMOS field using three independently
generated catalogs, and to compare the results with existing data.  We
also compare the amplitudes we measure to those found in a
semi-analytic model of galaxy formation.

The key feature of the COSMOS field is that it is completely covered
by the largest existing mosaic of image tiles from the Advanced Camera
for Surveys (ACS) on the Hubble Space Telescope
\citep{scoville_cosmos_hst_observations}. With respect to ground-based
surveys, in addition to the exceptional image quality, the great
advantage of this dataset is the superior photometric accuracy and
stability over the entire field of view of the survey, which in turn
makes it possible to measure the clustering of galaxies on large
scales and at faint magnitudes where the amplitude of the galaxy
correlation function $w(\theta)$ is very small.  \revised{The COSMOS
  field is currently the survey that probes the largest comoving
  scales at redshifts of around one. The total area covered, 2
  deg$^2$, does not exceed existing studies, in particular the UH8k
  study \citep[1.5 deg$^2$,][]{2003ApJ...585..191W} and the shallower
  DEEP2 measurements \citep[5.0 deg$^2$,][]{2004ApJ...617..765C}.
  However, COSMOS offers the unique combination of large contiguous
  area and depth, and also has the virtue of several independent and
  quite different imaging datasets in the same field.  }

We use this rich dataset to investigate the clustering properties of
the field galaxy population on degree scales.  In future papers, we
will present a more detailed study of galaxy clustering using
photometric redshifts \revised {which can be used, for example, to
  divide our galaxy catalogues by type and apparent magnitude}.  Our
objective here is simply to present the global properties of the field
in terms of simple two-point statistics for catalogs selected by
apparent magnitude. \revised{As our aim is to demonstrate the
  robustness of the results, we restrict ourselves to the $i$-band
  data, which is available for all three datasets considered here.}

The COSMOS field has been imaged by many ground-based facilities,
amongst them the Subaru telescope \citep{taniguchi_cosmos_subaru}, and
we compare the COSMOS ACS catalog (described in
\citealp{lea_cosmos_catalog}) with two ground-based catalogs: the
Subaru optical catalog described by \citet{capak_cosmos_catalog} and
the CFHTLS-T03 catalog used by McCracken et al. (2007, in preparation).
Each of these have significantly different masks, especially with
regard to ghosting around bright stars. As will be demonstrated, the
results of these independent determinations are in good agreement, and
display a consistently higher amplitude at faint magnitudes than has
been suggested in previous work.

\section{Catalogs and methods}

All of the three catalogs were prepared independently. For full
details of the ACS catalog, see \citet{lea_cosmos_catalog}; the
Subaru catalogs are described in \citet{capak_cosmos_catalog}. A
full description of the CFHTLS catalogs, based on the images
corresponding to the release CFHTLS-T03, can be found in McCracken
et al 2007.

The ACS data are constructed from a mosaic of 575 image tiles taken
over 588 orbits of the HST. The 50\% completeness limit of the
catalog is 26.6 F814W magnitudes.

SuprimeCam consists of 10 Lincoln-Labs 8k $\times$ 4k CCDs with a plate
scale of $0.2''$ per pixel.  However, a plate scale of $0.15''$ per
pixel was used for the final image to ensure that the images with good
seeing were not undersampled.  A special dither pattern including
camera rotations (Subaru is an alt-azimuth telescope) was used to
ensure every portion of the field was imaged by at least four different
CCDs. Further details can be found in \citet{taniguchi_cosmos_subaru}.
The Subaru catalog was based on a mosaic of 115 Subaru images taken
with the Suprime camera on the Subaru 8m telescope.  The image has a
median seeing of $0.6''$ and a $50\%$ completeness of $i=27.4$.

\revised{In order to produce the best possible catalogue for
  correlation function measurements, we re-extracted a catalogue
  ourselves from the Subaru tiles produced by Capak et al. We
  downloaded each tile and assembled them into a single large mosaic
  using the TERAPIX software tool \texttt{swarp}; we carried out the
  same procedure for the RMS maps. Following this, we used
  \texttt{sextractor} to extract a catalogue. Star-galaxy separation
  to $i=21.5$ was performed using the \texttt{flux\_radius} compactness
  parameter which measures the radius which encloses $50\%$ of an
  object's flux. Bright stars and defects were masked on the images. }

The CFHTLS catalog is derived from the TERAPIX CFHTLS-T03 release.
The CFHTLS stacks were taken using the Megacam camera on the 3.6m CFHT
telescope. Megacam covers 1 deg$^2$ with $0.205''$ pixels using 36
separate $2048\times4096$ Rockwell CCDs. Note that, unlike the Subaru
and ACS data, the CFHTLS-T03 image consists of a single Megacam
pointing. The
CFHTLS-T03 D2-i dataset comprises 153 images and has a median seeing
of $0.9''$.  The 50\% completeness of this dataset is 
$I_{\rm AB}\simeq 25.7$ magnitudes\footnote{For details see 
{\tt \url {http://terapix.iap.fr/cplt/tab\_t03ym.html}}}.

In all three catalogs, the star-galaxy separation is carried out
using a morphological classifier. We emphasize that the classifiers
were determined independently for each catalog; we did not, for
example, use the ACS morphologies to perform star-galaxy separation
on the CFHTLS or Subaru images, or use the Subaru images as
detection images for the ACS data. The Subaru images cover the full
2~deg$^2$ of the COSMOS field while the ACS tiling covers a total of
1.7~deg$^2$; the CFHTLS-T03 images cover just the central 1~deg$^2$
of the field. 

In each catalog, regions around bright stars and near the edges of the
field were masked. For the ACS catalogs, we use the same set of masks
that were used for weak lensing measurements
\citep{massey_cosmos_cosmicshear}. These masks also remove many
blended objects. 
\revised{After masking, in the Subaru catalogue there are 134,397
  galaxies in the magnitude range $20 < i < 25$; in the ACS catalogue
  and there are there are 124,665 galaxies in the same interval in
  magnitude. For the CFHTLS, there are 52,521 galaxies in the
  magnitude range $20 < i < 24$. The effective areas (total available
  area after masking) of the three surveys (Subaru, ACS and CFHT) is
  $1.6$, $1.5$, and $0.7$ deg$^2$ respectively
(i.e. completeness of 80\%, 88\% and 70\%). We have experimented with 
varying the degree of masking by `growing' the mask to eliminate
pixels adjacent to masked pixels. The results are robust even
when $>50\%$ of the area is masked.}

\begin{figure}
\epsscale{1.2}
\plotone{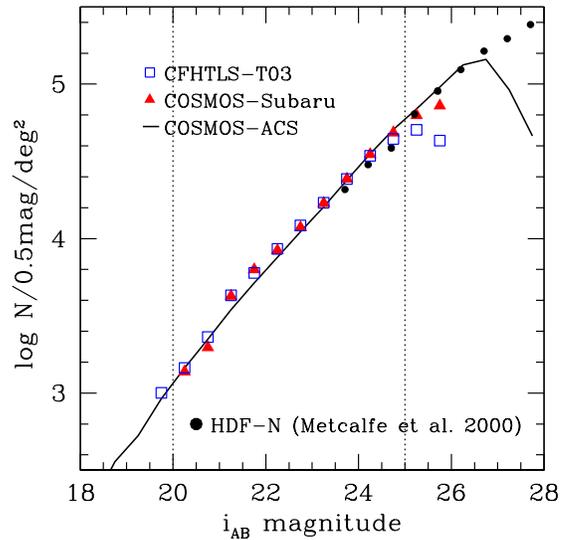}
\caption{Galaxy counts for the three catalogues presented in this
paper: ACS (solid line); CFHTLS-T03 (open squares) and Subaru (red
filled triangles). For reference, we also show galaxy counts extracted
from the HDF-N by \cite{2001MNRAS.323..795M}}
\label{fig:counts}
\end{figure}

\revised{Figure~\ref{fig:counts} shows the galaxy number counts extracted from the
  three catalogues. The dotted lines indicate the magnitude limits
  adopted in this paper. The slight `break' in the counts at $i*\simeq 21$ is
  an artefact caused by our morphologically based star-galaxy
  separation. The ACS count are slightly lower at $i\simeq 22$ as a
  consequence of the improved star-galaxy separation in this
  dataset. }

\section{Clustering measurements}

We selected galaxies in progressively fainter slices of apparent $i$
magnitude. For the purposes of the paper we assume that the
instrumental AB total magnitudes measured in each catalog are
equivalent; this is approximately true.  \citet{capak_cosmos_catalog}
present a detailed comparison of galaxy photometry between the three
catalogs described here. Figure 8 of their paper demonstrates that
total instrumental magnitudes in each catalog agree well, to
within 0.05 magnitudes.

\revised{For each slice, we measure $w$ for at range of angular separations
$\theta$ to $\theta+\delta \theta$ in a series of logarithmically
separated bins using the standard \citet{1993ApJ...412...64L} estimator, 
\begin{equation}
w ( \theta) ={\mbox{DD} - 2\mbox{DR} + \mbox{RR}\over \mbox{RR}}
\label{eq:1.ls}
\end{equation}
with the $DD$, $DR$ and $RR$ terms referring to the number of
data--data, data--random and random--random galaxy pairs between
$\theta$ and $\theta+\delta\theta$. }
\revised{The fitted amplitudes quoted in this paper assume a power-law
  slope for the galaxy correlation function,
  $w(\theta)=A_w(\theta/{\rm deg})^{-\delta}$; however this amplitude must be
  adjusted for the `integral constraint' correction, arising from
  the need to estimate the mean galaxy density from the sample
  itself. This can be estimated as \citep[e.g.][]{2005ApJ...619..697A},
\begin{equation}
C = {1 \over {\Omega^2}} \int\!\!\! \int w(\theta)\, d\Omega_1\, d\Omega_2,
\label{eq:5}
\end{equation}
where $\Omega$ is the area subtended by each of our survey fields. For
the COSMOS field, We find $C\sim 1A_w$ by numerical integration of
Equation~\ref{eq:5} over our field geometry and assuming that galaxies
closer than $1''$ cannot be distinguished.}

\begin{figure}
\epsscale{1.2}
\plotone{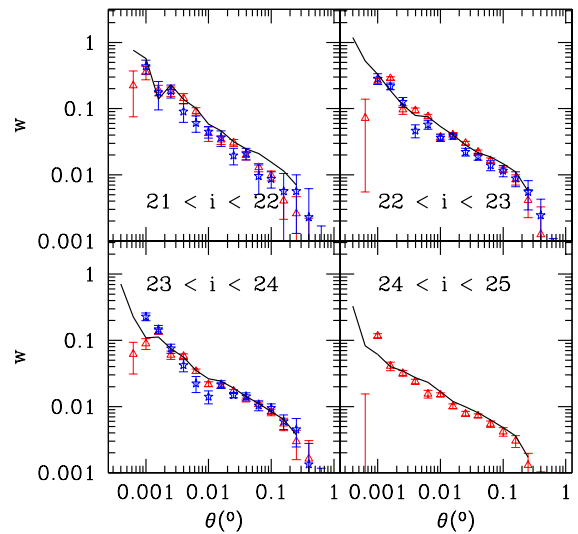}
\caption{The angular correlation function, $w(\theta)$ as a function of
  angular separation, $\theta$ in four slices of apparent
  magnitude. In each panel we show three different measurements from
  three different catalogs: ACS (connected lines); Subaru (triangles);
  CFHTLS/Megacam (stars).}
\label{fig:wtheta-comp}
\end{figure}

We used a sorted linked list in order to reduce computing time given
the very large number of objects in each slice. These results are
compared in Figure~\ref{fig:wtheta-comp}. The solid line shows
measurements from the ACS; the triangles and stars correspond to
measurements from Subaru and CFHT. At each angular bin for each
survey, the error bars plotted are simple bootstrap errors. Although
these are not in general a perfect substitute for a full estimate of
cosmic variance (e.g. using an ensemble of simulations), they should
give the correct magnitude of the uncertainty
\citep{1992ApJ...392..452M}.  \revised{In particular, these authors
  show that bootstrap errors yield sensible uncertainties on power-law
  fits to correlation-function data when the points are treated as
  independent.}

\begin{figure}
\epsscale{1.2}
\plotone{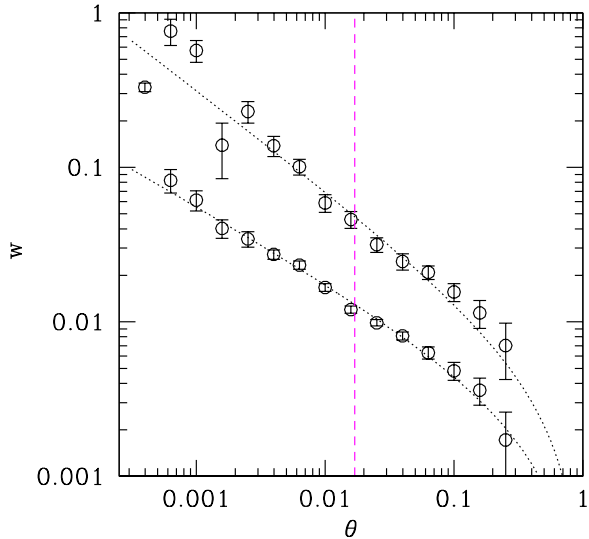}
\caption{The amplitude of the angular correlation function $w$, as a
  function of angular separation, $\theta$, for the ACS catalog.
  Measurements for galaxies selected in the magnitude ranges $21<i<22$
  and $24<i<25$ and are presented.  The dotted lines show the
  best-fitting power-law correlations (with slopes $-0.59$ and
  $-0.47$, respectively) and the
  integral correction for the COSMOS field  included. The dashed line
  is plotted at $1'$. There seems little evidence here for any deviation
  from power-law correlations, although the brighter bin does hint at an
  inflection around 1 arcmin. The correlation function for the fainter bin
  is clearly flatter.
}
\label{fig:wtheta-acs}
\end{figure}

For each of the four slices in apparent magnitude, the amplitude of
$w(\theta)$ measured in the Subaru data agrees well with the
measurements in the ACS. At very small angular separations the ACS data
are higher than the ground-based results. For the two faintest bins
($23 < i < 24$ and $24 < i < 25$) the agreement between the ACS
measurements and the Subaru measurements is excellent.

We now examine more closely the galaxy correlation function measured
from the ACS catalogs. As we have already seen, thanks to the
excellent resolution of the ACS images, we are able to measure
clustering amplitudes to small separations, on the order of $1''$.
Conversely, as a consequence of the large areal coverage of the ACS
COSMOS field, we can also measure amplitudes to large angular
separations. In Figure~\ref{fig:wtheta-acs} we show the angular
correlation function $w(\theta)$ as a function of angular separation
for bright and faint samples with $21<i<22$ and $24<i<25$
respectively. 
\revised{The dotted line shows the best-fitting lines
  with an integral constraint correction applied. For the bright bin,
  we find a best fitting slope of $-0.59\pm0.05$; for the fainter bin,
  $-0.47\pm0.02.$ We find unambiguously that in this magnitude limited
  sample, the galaxy correlation function becomes flatter towards
  fainter magnitude bins, in agreement with previous works
  \citep{2001A&A...376..756M}.}

\section{Comparisons with simulations}
\label{sec:repr-cosm-field}

\begin{figure}
\epsscale{1.2}
\plotone{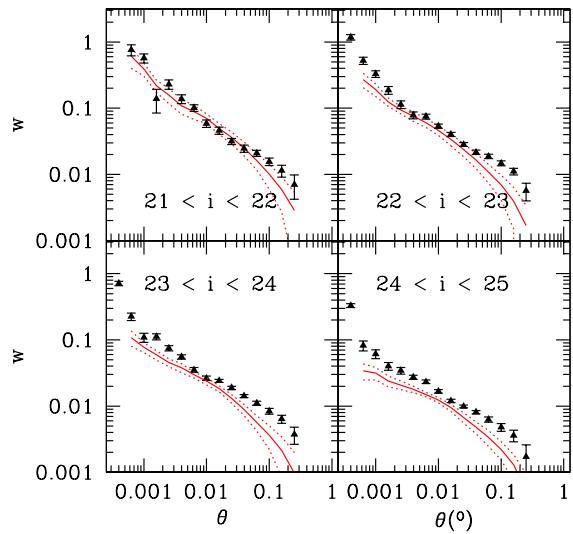}
\caption{The amplitude of the angular correlation function $w$, as a
  function of angular separation, $\theta$, for the ACS catalog. The
  lines show measurements made using mock catalogs extracted from the
  millennium simulation. The dotted lines show the one sigma scatter of
  the results for individual mock catalogs around their mean computed
  from the variance over twenty mock catalogs.  Note that these mock
  results automatically incorporate integral corrections due to the
  finite field size.  }
\label{fig:wtheta-millennium}
\end{figure}

In the previous Sections we established that measurements between the
different datasets are consistent. In this Section, we compare our
measurements to those made on catalogs produced by
\cite{2006astro.ph..9636K}. These catalogs were created using a
semianalytic model to simulate galaxy formation within the evolving
halo population of the extremely large Millennium Simulation. These
simulated Universes were then `observed' to produce light-cones, which
can be then used to produce observations of identical geometry to those
of real catalogs.  \citep{2005MNRAS.360..159B,2006astro.ph..9636K}.
Each COSMOS light-cone covers 2~deg$^2$.
Figure~\ref{fig:wtheta-millennium} shows, as before, the amplitude of
$w(\theta)$ as a function of angular separation in four magnitude
slices. Points with error bars show measurements from the ACS data. The
solid line shows the average of measurements made from twenty
light-cones extracted from the millennium simulation; the dotted lines
show the amplitude of the $\pm1 \sigma$ error bars.

For the two brighter slices, $21<i<22$ and $22<i<23$, the agreement
between the simulations and observations at intermediate and small
scales is remarkably good. At fainter magnitudes, and at larger scales,
the amplitudes measured in the COSMOS catalogs are consistently
higher than the prediction of the simulations. Since the simulations
allow us to assess the amplitude of cosmic variance directly, we
can easily conclude that this discrepancy cannot plausibly be attributed
to the COSMOS field having above-average clustering. It therefore
appears that the model predictions for $w(\theta)$ at $i\simeq 24$
are too low; we discuss this further below.

\section{Comparisons with literature measurements}

From our measurements it is clear that the form of $w(\theta)$ in the
COSMOS field does not correspond to a simple power law with a slope
that is independent of median magnitude of the sample. In the past,
determinations of galaxy clustering amplitude were usually given at a
fixed angular scale over a small range of angular separations. From
our data, it is clear that fitting $w(\theta)$ only at small angular
separations will result in different fitted amplitudes as compared to
a fit over the entire range. Therefore, in order to compare with
results presented in the literature, we choose to carry out fits over
\revised{a similar range of angular separations}.

\begin{figure}
\epsscale{1.2}
\plotone{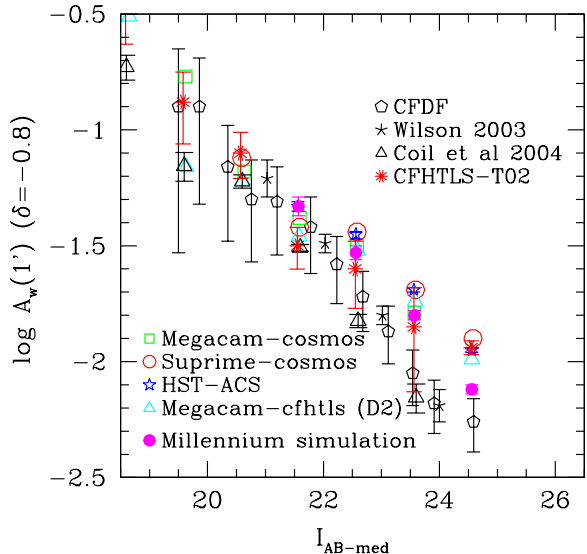}
\caption{Comparisons of the COSMOS results for the dependence of
the amplitude of $w(\theta)$ on depth, compared with existing
measurements and with the mock results. We choose to specify
the amplitude at 1 arcmin, based on a $\delta=-0.8$ power-law
fit around this point. The results are highly insensitive to this choice
of slope.}
\label{fig:wtheta-comparisons}
\end{figure}
\nocite{2003ApJ...585..191W}

In Figure~\ref{fig:wtheta-comparisons} we show the fitted amplitudes of
$w(\theta)$ as a function of the median magnitude of each slice. In
comparing with literature measurements, we see that at bright
magnitudes $(i\simeq 20)$ our measurements are approximately in
agreement with the values presented in the literature. However, by
$i\simeq 23$, the COSMOS field measurements are significantly higher
than, for example the CFDF measurements of \citet{2001A&A...376..756M}.
This is true at all angles and for all measurements, whether they are
from the ground based or space-based catalogs.

Photometric redshifts are available for all objects in the COSMOS field
\citep{mobasher_cosmos_photz}. The slice with $22<i<23$ has a median
redshift of $z\simeq0.8$; the next faintest slice is at $z\simeq0.9$.
If we examine Figure 2 of \citet{scoville_cosmos_lss} we can see there
is a significant over-density in the redshift range $0.7<z<0.9$. It is
possible that this structure could contribute to the enhanced signal on
large scales seen in our data in these slices.

Some independent evidence exists that this structure raises the
COSMOS clustering amplitude above the ensemble average. In
Figure~\ref{fig:wtheta-comparisons} we have plotted the average
amplitude of $w$ as a function of apparent median magnitude for the
four independent deep fields of the Canada-France Legacy Survey (one
of which is the D2 field described above), totalling an effective
area of 3.2 deg$^2$. Error bars correspond to the variance over all
four fields. Interestingly, at bright magnitudes, the CFHTLS
magnitudes agree with the COSMOS measurements, and other literature
values; at intermediate magnitudes ($22<i<23$) they are between the
COSMOS values and those from other surveys, whereas at fainter
magnitudes they agree perfectly with the COSMOS measurements,
presumably because the median redshift probed by both surveys at
these magnitude limits is beyond that of the rich structure in the
COSMOS field. The size of the cosmic variance error bars on the
CFHTLS measurements are also largest at intermediate magnitudes.
These measurements are discussed in greater detail in a forthcoming
paper (McCracken et al., in preparation).

\section {Discussion and conclusions}

This paper has presented measurements of the angular two-point
correlation function in the COSMOS field, $w(\theta)$, and how it
depends on $i$-band magnitude depth. We have shown that consistent
results are obtained using three independent datasets: HST ACS; Subaru
SuprimeCam; and CFHT MegaCam. The results agree well at bright
magnitudes ($i\simeq 22$) with previous measurements and with the
predictions of semi-analytic mock catalogs constructed from the
Millennium Simulation (MS). The only slight caveat here is that the
results at $\theta > 10$ arcmin are at the high end of the MS
predictions, which may reflect a single rich $z\simeq 0.8$ structure
in the field.

At fainter magnitudes, however, a different picture emerges.  By the
time we reach the $24<i<25$ bin, the COSMOS measurements are
consistently a factor 2 higher than the MS predictions at all angles.
Moreover, the COSMOS measurements are consistent with the four-field
average of measurements from the CFHTLS survey.  This discrepancy is
well beyond the compass of cosmic variance from limited numbers of rich
structures, as measured via the ensemble of simulations.  Thus, barring
some undetected systematic that is consistent between all the datasets
we have used, the conclusion must be that the MS predictions are too
low at these magnitude levels.  This could arise in a number of ways:
the predicted degree of bias at high redshifts might be too low; 
the luminosity function might be incorrect, resulting in too high a predicted
mean redshift at these depths; or alternatively the MS may miss
foreground pairs of intrinsically faint galaxies because of the
resolution limit of the simulation. The first possibility is
particularly interesting given the current debate over the
normalization of the primordial power spectrum, $\sigma_8$. The MS used
$\sigma_8=0.9$, whereas WMAP favors a smaller result -- perhaps as low
as $\sigma_8=0.7$ (Spergel et al. 2006). Since high-redshift galaxies
are strongly biased already, a reduced $\sigma_8$ will in fact boost
the predicted galaxy clustering (for a given galaxy mass). These issues
will be explored further in future papers, where we make direct use of
the photometric redshift data in the COSMOS field.

\acknowledgements 
This work is based in part on data products
produced at TERAPIX located at the Institut d'Astrophysique de
Paris. H. J. McCracken wishes to acknowledge the use of TERAPIX
computing facilities.  

The HST COSMOS Treasury program was supported through NASA grant
HST-GO-09822. We wish to thank Tony Roman, Denise Taylor, and David
Soderblom for their assistance in planning and scheduling of the
extensive COSMOS observations. We gratefully acknowledge the
contributions of the entire COSMOS collaboration consisting of more
than 70 scientists. More information on the COSMOS survey is available
at \url{http://cosmos.astro.caltech.edu}. It is a
pleasure to acknowledge the excellent services provided by the NASA
IPAC/IRSA staff (Anastasia Laity, Anastasia Alexov, Bruce Berriman and
John Good) in providing online archive and server capabilities for the
COSMOS datasets.  Data for the Millennium Simulation are publically
available at \url{http://www.mpa-garching.mpg.de/millennium}.

 \clearpage


\end{document}